\documentclass[12pt,preprint]{aastex}

\usepackage{graphicx}
\usepackage{natbib}
\shorttitle{SMA CO Observations of SVS\,13}%
\shortauthors{X.~Chen et al.}%

\begin{document}

\title{Rotating Bullets from A Variable Protostar}

\author{Xuepeng~Chen\altaffilmark{1}, H\'{e}ctor~G.~Arce\altaffilmark{2}, Qizhou~Zhang\altaffilmark{3}, Ralf Launhardt\altaffilmark{4}, and Thomas Henning\altaffilmark{4}}

\affil{$^1$Purple Mountain Observatory, Chinese Academy of Sciences, 2 West Beijing Road, Nanjing 210008, China; xpchen@pmo.ac.cn}%
\affil{$^2$Department of Astronomy, Yale University, Box 208101, New Haven, CT 06520-8101, USA}%
\affil{$^3$Harvard-Smithsonian Center for Astrophysics, 60 Garden Street., Cambridge, MA 02138, USA}%
\affil{$^4$Max Planck Institute for Astronomy, K\"{o}nigstuhl 17, D-69117 Heidelberg, Germany}

\begin{abstract}

We present SMA CO\,(2--1) observations toward the protostellar jet driven by SVS\,13\,A, a variable protostar in the NGC\,1333 star-forming
region. The SMA CO\,(2--1) images show an extremely high-velocity jet composed of a series of molecular ``bullets". Based on the SMA CO 
observations, we discover clear and large systematic velocity gradients, perpendicular to the jet axis, in the blueshifted and redshifted bullets. 
After discussing several alternative interpretations, such as twin-jets, jet precession, warped disk, and internal helical shock, we suggest that 
the systematic velocity gradients observed in the bullets result from the rotation of the SVS\,13\,A jet. From the SMA CO images, the measured 
rotation velocities are 11.7--13.7\,km\,s$^{-1}$ for the blueshifted bullet and 4.7\,$\pm$\,0.5\,km\,s$^{-1}$ for the redshifted bullet. The estimated 
specific angular momenta of the two bullets are comparable to those of dense cores, about 10 times larger than those of protostellar envelopes, 
and about 20 times larger than those of circumstellar disks. If the velocity gradients are due to the rotation of the SVS\,13\,A jet, the significant 
amount of specific angular momenta of the bullets indicates that the rotation of jets/outflows is a key mechanism to resolve the so-called ``angular 
momentum problem" in the field of star formation. The kinematics of the bullets suggests that the jet launching footprint on the disk has a radius 
of $\sim$\,7.2--7.7\,au, which appears to support the extended disk-wind model. We note that further observations are needed to comprehensively 
understand the kinematics of the SVS\,13\,A jet, in order to confirm the rotation nature of the bullets.

\end{abstract}


\clearpage

\section{INTRODUCTION}

Stars form from the gravitational collapse of dense cores in molecular clouds. It has been long recognized that
in the star formation process specific angular momenta need to be reduced by 6--7 orders of magnitude from a 
dense core to a typical main-sequence star, like the Sun. This puzzle has been regarded as the ``angular 
momentum problem" in the field of star formation (e.g., Spitzer 1978; Bodenheimer 1995). Obviously, the 
extraction of excess angular momentum is essential in the star formation process.

In the early phase of star formation, when a young stellar object (YSO) accretes mass from its surrounding disk, 
it also ejects mass in a bipolar jet. The ejected material can accelerate entrained gas to velocities larger than those of 
the cloud, thereby creating a molecular outflow. Numerous observations indicate that most, if not all, stars drive 
accretion-powered jets and outflows during their formation (Reipurth \& Bally 2001; Arce et al. 2007). The jets are 
believed to be driven by rotating disks through magneto-centrifugal processes, although the precise origin of jets 
from YSOs is still in debate (Blandford \& Payne 1982; Pudritz et al. 2007; Shang et al. 2007). The rotation of jets 
and outflows about their symmetry axis is widely considered to be a key mechanism to remove angular momentum 
from forming protostars, and thus, at least partially, to solve the angular momentum problem in star formation.
Furthermore, rotational kinematics of jets and outflows can be used to study jet-launching mechanism and to discriminate 
different jet-launching theories (Frank et al. 2014).

As a consequence, many groups have searched for rotation signatures in jets and outflows, i.e., radial velocity 
gradients or shifts across the axes of jets/outflows. However, the detection of jet/outflow rotation is extremely 
difficult, as both high angular and spectral resolution are required (see, e.g., Ray et al. 2007). As shown by Pesenti 
et al. (2004), the underlying rotation signature of jets/outflows could be easily diluted and made undetectable in 
low angular resolution observations. During the past two decades, radial velocity gradients/shifts have been observed 
in a small number of jets and molecular outflows (see reviews by Bachiotti et al. 2009; Belloche 2013; Frank et al. 
2014). However, several alternative interpretations other than rotation have been suggested to explain the observed 
velocity gradients/shifts across jets/outflows, such as asymmetric shocking against a warped disk (Soker 2005), jet 
precession (e.g., Cerqueira et al. 2006), helical MHD shocks (Fendt 2011), and twin-jets (Soker \& Mcley 2013). 
Hence, one must be cautious when interpreting observed velocity gradients across jet/outflows. As suggested by 
Belloche (2013), the following requirements are needed to convincingly detect rotation in jets or outflows. First of all, 
the kinematic signature, a velocity gradient perpendicular to the jet axis, has to be consistent all along the jet length. 
Second, the signature should be consistent between both lobes and with the disk rotation. Finally, the velocity profile 
should be spectrally resolved. Taking these requirements into account reduces the number of candidate rotating 
jets/outflows to less than 10. 

A widely accepted prototypical example of such rotational jets/outflows is DG~Tau, a small jet driven by a Class\,II
T~Tauri star. High angular resolution HST observations revealed in the DG~Tau jet a large velocity shift across 
the jet axis ($\sim$\,20\,km\,s$^{-1}$), which is consistent with the rotation direction of the circumstellar disk (Bachiotti 
et al. 2002). Another promising candidate is the molecular outflow in CB\,26, driven by an isolated T~Tauri star 
(Launhardt et al. 2009). As calculated by Launhardt et al. (2009), the (specific) angular momentum in the CB\,26 
circumstellar disk can be extracted by the outflow in a time scale of $\sim$\,10$^6$ years. A few other rotational 
jet/outflow candidates, e.g., TH\,28 and CW~Tau (Coffey et al. 2007), show similar rotational signatures in the high 
angular resolution observations. These jets/outflows are so far the prime candidates for studying jet-driving mechanisms.

Although promising rotation jets/outflow candidates from a few T~Tauri stars have been observed, these stellar objects 
are well past their main accretion/ejection phase. As most of the stellar mass is accreted during the protostellar phases,
most of the angular momentum loss must have occurred by then as well, with only a small fraction left to deal with during 
the T~Tauri evolution. In particular, the initial amount of angular momentum, and its evolution and re-distribution appear to 
be key factors in many of the processes which determine the protostellar fragmentation, the final stellar mass(es), and the 
existence and morphology of planetary systems.
Therefore, it is of prime importance to detect and study rotating jets/outflows from protostars when the removal/evolution 
of angular momentum is more crucial. In recent years, a few rotational outflow candidates were found in embedded 
protostars, although none of them have been confirmed (see, e.g., Belloche 2013).
In a series of studies toward HH\,211, Lee et al. (2009) reported a tentative detection of rotation ($\sim$\,1.5\,km\,s$^{-1}$
velocity shift), but the rotation direction is opposite to the gradient direction found by Lee et al. (2007). For the HH\,212 outflow, 
the tentative detection of a rotation velocity shift ($\sim$\,1.5\,km\,s$^{-1}$) reported by Lee et al. (2008) is not confirmed by 
other observations (e.g., Codella et al. 2007; Correia et al. 2009). 
Zapata et al. (2010b) reported a candidate rotating jet in Ori-S6. However, the driving source of the jet is a close binary 
system with a separation of $\sim$\,50\,au (Zapata et al. 2007), implying that the rotation signature may be explained 
by the twin-jets scenario. In addition, the referred launching footprint of the Ori-S6 jet, $\sim$\,140/(sin\,$i$)$^{3/4}$\,au 
(where $i$ is the inclination angle of the jet; see Zapata et al. 2010b), is larger than that of the binary separation, not to 
mention the circumstellar disk in the system. Furthermore, the direction of the velocity gradient across the jet is also opposite 
to that of the gradient found in the circumbinary ring (Zapata et al. 2010a).
The other two protostellar outflows, NGC\,1333 IRAS\,4A (Choi et al. 2011) and IC\,348\,MM2 (Pech et al. 2012), may be also 
explained by the twin-jets scenario (see Soker \& Mcley 2013 and Rodr\'{i}guez et al. 2014, respectively). Therefore, there is 
no convincing detection of rotating jets/outflows from protostars yet.

In this paper, we report Submillimeter Array\footnote{The Submillimeter Array is a joint project between the 
Smithsonian Astrophysical Observatory and the Academia Sinica Institute of Astronomy and Astrophysics 
and is funded by the Smithsonian Institution and the Academia Sinica.} (SMA; Ho et al. 2004) CO\,(2--1) line
observations toward the protostellar jet driven by SVS\,13\,A, a Class\,0/I transition protostar located in the 
Perseus NGC\,1333 star-forming region (Bachiller et al. 2000; Chen et al. 2009). The distance of SVS\,13\,A 
is accurately determined to be 235\,$\pm$\,18\,pc based on VLBI parallax measurements of masers 
associated with this object (Hirota et al. 2008). SVS\,13\,A is well-known for driving an extremely high-velocity 
CO outflow (see, e.g., Bachiller et al. 1990; Masson et al. 1990), which records the highest velocity CO outflow 
known thus far.
With Very Large Array (VLA) 3.6\,cm observations, Anglada et al. (2000) resolved source SVS\,13\,A into a close 
binary system, VLA\,4A and VLA\,4B, with an angular separation of 0\farcs3. Further VLA 7\,mm observations indicated 
that only VLA\,4B is associated with a circumstellar disk (Anglada et al. 2004). Indeed, high angular resolution 
near-infrared observations at the Keck telescopes show that VLA\,4B is the driving source of the SVS\,13\,A protostellar 
jet (Hodapp \& Chini 2014).

Another interesting characteristic of SVS\,13\,A is its variability. After its discovery in infrared observations (Strom et al.
1976), this source experienced a sudden increase in brightness around 1990 ($\sim$\,1.5 magnitude changed in the 
$K$ band; see Hodapp \& Chini 2014 and references therein). This phenomenon is traditionally classified as either
FU Orionis (FUor) or EX Lupi (EXor) type outbursts, depending on the duration of the outburst and its spectrum (see
a recent review by Audard et al. 2014). In both classes, an increase in the accretion rate onto the star is thought to be 
responsible for the luminosity increase.
After analyzing the infrared data collected over the course of the past 24 years, Hodapp \& Chini (2014) concluded that 
the elevated post-outburst level of accretion and, therefore, jet activity in SVS\,13\,A has persisted for the past two decades.
The light curve of SVS\,13\,A shares the long duration maximum with FUors, while spectroscopically, it resembles an EXor, 
and the small outburst amplitude resembles neither (see Hodapp \& Chini 2014). Therefore, SVS\,13\,A represents an 
object somewhere between those two classical cases.

There is growing evidence that protostellar accretion cannot be steady in general, but must alternate between high and low 
states of accretion (Hartmann \& Kenyon 1996; Audard et al. 2014). Both observational and theoretical studies suggest 
that variable protostellar accretion with episodic outbursts may be a standard phenomenon in star formation. A recent discovery 
of an outbursting Class\,0 protostar indicates that episodic accretion occurs already at the very early phase of star formation 
(Safron et al. 2015), which is consistent with the prediction of theoretical simulations (e.g., Vorobyov \& Basu 2015).  
On the other hand, episodic ejection events seem also to be a common property of protostellar jets and outflows (Reipurth \& 
Bally 2001; Arce et al. 2007). Due to the physical link between accretion and ejection, this could be explained by variations 
in the accretion rate onto the forming star, which results in variations on the velocity of the ejected matter, hence the creation 
of a series of shocks. The SVS\,13\,A jet, also well-known as Herbig-Haro (HH) flow 7-11, comprised of a long chain of 
individual shock fronts (see, e.g., Raga et al. 2013) and molecular outflow clumps (see, e.g., Plunkett et al. 2013), is a typical 
example of such protostellar jets. Therefore, the variability of SVS\,13\,A and its jet structure also provide an important 
opportunity to study the relationship between episodic accretion and outbursts.

\section{OBSERVATIONS AND DATA REDUCTION}

The data presented in this work were taken with the Submillimeter Array (Ho et al. 2004), an interferometer array
of eight 6-meter antennas at Mauna Kea, Hawaii. We observed SVS\,13 at 230\,GHz on 2008 December 7th in the 
SMA compact configuration. Eight antennas were used in the array, providing baselines with projected lengths from 
$\sim$\,9 to 59\,k$\lambda$.
The receiver was set up to cover the frequency ranges of 219.5$-$221.3 GHz and 229.5$-$231.3 GHz in the lower 
and upper sidebands, respectively. This setup includes three isotopic CO lines of $^{12}$CO\,(2--1) (230.538\,GHz), 
$^{13}$CO\,(2--1) (220.399\,GHz), and C$^{18}$O\,(2--1) (219.560\,GHz), as well as the 1.3\,mm dust continuum 
emission. In the observations, system temperatures of SVS\,13 ranged from 130 to 290\,K (depending on elevation), 
with a typical value of $\sim$\,160\,K. The SMA primary beam is approximately 55$''$ at 230\,GHz. 

The data were reduced using the IDL MIR package (Qi 2005). During the reduction, visibility points with clearly deviating 
phases and/or amplitudes were flagged. The passband (spectral response) was calibrated through observations of the
strong quasar 3C273. Time variations in amplitude and phase were calibrated through frequent observations of quasars 
0238+166 and 0359+509. Uranus was used for absolute flux calibration, and we estimate a flux accuracy of about 15\%. 

The calibrated visibility data were then imaged using the MIRIAD toolbox (Sault et al. 1995). The data were imaged by first 
computing the inverse Fourier transform of the data with the {\it invert} task. The dirty map was then deconvolved using the 
CLEAN algorithm with the {\it clean} task down to 1--2 times the rms noises level. The cleaned components were then 
convolved with the clean beam using the {\it restore} task. With robust {\it uv} weighting +1, cleaned maps of the CO\,(2--1) 
emission were produced with an effective synthesized beam size of 2\farcs8\,$\times$\,2\farcs6. 
The 1\,$\sigma$ rms noise level is $\sim$\,30--35\,mJy\,beam$^{-1}$ at a velocity resolution of $\sim$\,1.0\,km\,s$^{-1}$.

\section{RESULTS: SMA CO\,(2--1) MAPS}

In the SMA observations, the CO\,(2--1) line emission from SVS\,13\,A is detected in a velocity range from approximately 
$-$150 to +150\,km\,s$^{-1}$, while the systemic LSR radial velocity of the ambient molecular cloud is about +9\,km\,s$^{-1}$ 
(see Chen et al. 2009). Such a large velocity range of CO emission represents so far the highest velocity molecular outflow 
detected from a protostar. 
Following Bachiller et al. (2000), we have divided the high-velocity emission into a limited number of velocity intervals. In
this work we focus on the ``extremely high velocity" (EHV) components (radial velocities more than 70\,km\,s$^{-1}$ over 
the cloud systemic velocity), whereas other velocity components will be studied in another work. 

Figure~1 shows the velocity-integrated intensity map of the EHV CO\,(2--1) emission from SVS\,13\,A, plotted on the SMA 
1.3\,mm dust continuum image. The CO map shows a high-velocity blueshifted jet, emanating from SVS\,13\,A towards the
southeast. The position angle of the jet is 154\,$\pm$\,1 degree (measured east from north). This jet consists of three well 
aligned knots, or so-called ``molecular bullets". The LSR velocities of the three blueshifted bullets are estimated to be $-$70, 
$-$110, and $-$150\,km\,s$^{-1}$, respectively; the three bullets are named blue bullets 1, 2, and 3 (see Figure~1). The 
projected distance between SVS\,13\,A and blue bullet 3, the bullet that is the farthest from the driving source, is approximately 
22$''$ ($\sim$\,5200\,au at the distance of 235\,pc). To the northwest of SVS\,13\,A and roughly aligned with the blueshifted jet, 
the EHV redshifted emission shows another bullet, with an LSR velocity of about +145\,km\,s$^{-1}$. This redshifted bullet,
referred to as red bullet in this work, is $\sim$\,16$''$ (or $\sim$\,3800\,au) away from SVS\,13\,A, spatially coincident with the 
small redshifted clump ``R" found by Bachiller et al. (2000). 

Figure~2 shows the intensity-weighted velocity map of the CO\,(2--1) emission (1st moment map). The images reveal in blue 
bullet 2 a previously undiscovered systematic velocity gradient, which is perpendicular to the local jet axis (defined by the jet 
wiggle morphology, see below). A systematic velocity gradient is also revealed in the red bullet, and the direction of the gradient 
is consistent with that of blue bullet 2, i.e., the left side of the jet axis is blueshifted while the right side is redshifted. The CO velocity 
channel maps of the blue bullets and red bullet are shown in Appendix (Figures~7 to 9), where detailed kinematic information 
of the bullets can also be found.
On the other hand, as seen in Figure~2, the velocity field of blue bullet 1 shows a linear gradient along the jet axis, increasing 
from the driving source SVS\,13\,A to the southeast. This feature, generally referred to as the ``Hubble-law", is frequently seen 
in jet-driven outflows (Lada \& Fich 1996; Arce \& Goodman 2001). It appears that the velocity field of the red bullet is also partially
affected by this ``Hubble-law" effect in the direction of the jet. The velocity field of blue bullet 3 also shows this ``Hubble-law"
kinematics (see Figure~2), which will not be further discussed in this work.

Figure~3 shows the position-velocity (PV) diagrams for the two cuts across blue bullet 2 and red bullet (see Figure~2).
The PV diagrams also indicate the presence of systematic velocity gradients perpendicular to the jet axis in the two bullets. 
For blue bullet 2, the velocity shifts ($v_{\rm shift}$) and shift distances ($L_{\rm shift}$), measured between the blue and 
red peaks in the PV diagram (see Fig.\,3), are $\sim$\,15.0--17.6\,km\,s$^{-1}$ and $\sim$\,3\farcs7--4\farcs5 (depending 
on cut positions), respectively. For the red bullet, the velocity shift and shift distance are $\sim$\,5.6\,$\pm$\,0.5\,km\,s$^{-1}$ 
and $\sim$\,10\farcs3\,$\pm$\,0\farcs5, respectively. 
At the distance of 235\,pc, the derived velocity gradients are $\sim$\,2.5--4.1\,$\times$\,10$^3$\,km\,s$^{-1}$\,pc$^{-1}$ for 
blue bullet 2 and $\sim$\,0.5\,$\times$\,10$^3$\,km\,s$^{-1}$\,pc$^{-1}$ for the red bullet. These gradients are about two 
orders of magnitudes larger than those typically observed in protostellar envelopes (e.g., Belloche et al. 2002; Chen et al. 2007; 
Tobin et al. 2011).

\section{DISCUSSION}

\subsection{Episodic Ejections of Bullets}

As shown in Figure~4, the blueshifted SVS\,13\,A jet presents a wiggling morphology, which is frequently seen in YSOs jets/outflows,
e.g., HH\,30 (Anglada et al. 2007) and HH\,211 (Lee et al. 2010). This wiggle of the jet is widely thought to be caused by the orbital 
motion of the jet source in a binary system (see, e.g., Fendt \& Zinnecker 1998; Masciadri \& Raga 2002). We therefore model the 
wiggle of the jet using a similar method as described in Lee et al. (2010). In this binary model, the orbit has a period $P_{\rm o}$ and 
a radius $R_{\rm o}$ (the orbital radius of the jet source with respect to the binary center of mass); the ejection velocity of the jet has 
a component in the orbital plane due to orbital motion, $v_{\rm o}$, and a component perpendicular to this orbital plane $v_{\rm j}$; 
the periodic length is defined as $\Delta$$\it z$~$\equiv$~$v_{\rm j}$$P_{\rm o}$ and the velocity ratio as 
$\kappa$~$\equiv$~$v_{\rm o}$/$v_{\rm j}$.

Figure~4 shows the best fit of this model to the blueshifted jet. As seen in Fig.\,4, the wiggle of the jet can be reasonably fitted by 
this model. This agreement between the observations and model appears to support the scenario that the wiggle of the blueshifted 
jet results from the orbital motion of the driving source. The best-fit parameters are periodic length $\Delta$$\it z$ = 8\farcs5\,$\pm$\,0\farcs2 and 
$\kappa$ = 0.075\,$\pm$\,0.003. The parameters of the model are directly related to the observations: $\kappa$ is related to the 
half-opening angle $\alpha$ of the jet cone, $\kappa$ = $\tan \alpha$~$\sin i$, where $i$ is the inclination angle with respect to the line 
of sight and is adopted as 40$^\circ$ in this work (see, e.g., Davis et al. 2001; Takami et al. 2006). The orbital radius $R_{\rm o}$ is 
related to the periodic length $\Delta$$\it z$ of the wiggle,

\begin{equation}
R_{\rm o} =  \frac{\Delta z\,\tan \alpha}{2\pi}.
\end{equation}

\noindent The derived orbital radius $R_{\rm o}$ is approximately 0\farcs16\,$\pm$\,0\farcs02, which is consistent with the results 
from the VLA observations (Anglada et al. 2000;  2004). 

However, the period of the curve in Figure~4 showing the best-fit model does not precisely match the spacings between the blue peaks. 
In the SMA CO observations, the spacing between bullets 1 \& 2 ($\sim$\,2600\,au) is different from that between bullets 2 \& 3 
($\sim$\,2000\,au). The proper motions of the SVS\,13\,A jet have been well studied at optical and near-infrared wavelengths in the 
past two decades (see, e.g., Raga et al. 2013 and Hodapp \& Chini 2014, and references therein). These studies suggest that the 
proper motions of the jet range from $\sim$\,15\,km\,s$^{-1}$ (inner jet; Hodapp \& Chini 2014) to $\sim$\,40\,km\,s $^{-1}$ (outer 
large-scale jet; e.g., Khanzadyan et a. 2003). Therefore, there is a possibility that the varying proper motions of the jet cause the 
unequal spacings between the bullets, which in turn results in the difference between the model and observations.

Since jets are accretion-driven, jet properties that change with time and space can provide important constraints on past temporal 
variations in the ejection/accretion system (see Arce et al. 2007; Frank et al. 2014). As seen in the SMA CO images, the bullets in
the SVS\,13\,A jet, representing abrupt enhancements in velocity and density, are very likely the results of episodic ejections, as seen 
in other jets/outflows, e.g., HH\,80-81 (Qiu \& Zhang 2009), L1448C (Hirano et al. 2010), and HH\,46/47 (Arce et al. 2013). As introduced 
in Section\,I, SVS\,13\,A itself also displays a high variability, similar to a FUor-type object (Hodapp \& Chini 2014). Therefore, it is of 
interest to relate the episodic ejections of the bullets with the FUor-like outbursts of SVS\,13\,A. Based on the spacings between the 
blue bullets, the estimated period for the episodic ejections in the SVS\,13\,A jet is $\sim$\,240--830\,yr (depending on the proper motion 
velocities), which is consistent with the periods predicted for the FUor type outbursts (Audard et al. 2014). 

\subsection{Rotation of Bullets}

SMA CO\,(2--1) observations show clear systematic velocity gradients, perpendicular to the jet axis, in both the blue bullet 2 and red 
bullet in the SVS\,13\,A jet. The velocity gradients in the two bullets are roughly in the same direction. Although an obvious explanation 
is that the velocity gradients result from the rotation of the bullets, as introduced in Section\,I, several alternative interpretations for such 
systematic velocity gradients should also be considered, such as twin-jets configuration, jet precession, warped disk, and helical MHD 
shocks.

Based on an SMA dust continuum survey toward 33 Class\,0 protostars, Chen et al. (2013) found that about two-thirds of protostars are 
binary/multiple systems, with separations ranging from 5000 down to 50\,au. A higher angular resolution VLA survey further confirms a high 
multiplicity frequency in protostars with smaller separations ranging from 100 down to 10\,au (Tobin et al. 2016). Therefore, it is 
necessary to consider the (non-rotating and non spatially resolved) twin-jets model that could possibly lead to the observed velocity field
(see discussion in Launhardt et al. 2009). Indeed, Launhardt et al. (2009) and Soker \& Mcley (2013) found that the twin-jets scenario 
seems to apply to the cases of CB\,26 and NGC\,1333 IRAS\,4A, respectively. 
In SVS\,13\,A, VLA radio observations detected a close binary system (VLA\,4A and  VLA\,4B) with a separation of $\sim$\,0\farcs3 
(Anglada et al. 2000). However, higher angular resolution near-infrared observations ($\sim$\,0\farcs01 or $\sim$\,2.4\,au) at the 
Keck telescopes clearly show that the high-velocity SVS\,13 jet is driven only by source VLA\,4B and no jet is detected from source 
VLA\,4A (Hodapp \& Chini 2014). The Keck results are consistent with the VLA observations that only source VLA\,4B is associated 
with a circumstellar disk (and thus jet) in this binary system (Anglada et al. 2004). Therefore, both the Keck and VLA observations rule 
out the twin-jets configuration in the case of SVS\,13\,A.

Another potential explanation is that the SVS\,13\,A jet is a precessing (non-rotating) jet. The precession of the jet is driven by tidal 
interactions between the disk from which the jet is launched and a companion star on a non-coplanar orbit. In this scenario, the 
velocity gradient perpendicular to the jet axis is a result of outflow gas being entrained along different directions at different times.
However, precessing jets generally produce point-symmetric (S-shaped) wiggles (see, e.g., L1157, Takami et al. 2011). If the 
velocity gradients in blue bullet 2 and the red bullet were indeed caused by a precessing jet, the directions of the velocity gradients in 
the two bullets would be in the opposite direction, but in the observations the velocity gradients of the two bullets are approximately 
in the same direction. Therefore, we consider that the precessing jet explanation does not apply to the case of SVS\,13\,A. 
Moreover, we note that it is still uncertain whether precession is able to produce velocity gradients mimicking rotation signatures in 
a jet. For example, three-dimensional molecular line simulations in Smith \& Rosen (2007) found no signature for rotation in precessing 
jets, which is in contrast to the results of Cerqueira et al. (2006), who found velocity gradients in both rotating and precessing jets.

Suggested by Soker (2005), the interaction of a jet with a twisted-tilted (warped) disk may also produce observed asymmetry velocity 
profiles. In this scenario, there is an assumed inclination between the jet and the outer parts of the disk. Namely, the inner disk is 
perpendicular to the jet, while the outer disk flares in a point-symmetric manner: one side up and the opposite (relative to the central 
driving source) down. Another assumption in this scenario is that the jet interacts with the ambient gas above the disk surface and is 
more slowed down on the warped side of the disk than on the opposite side. 
Nevertheless, high angular resolution VLA observations show no hints for the existence of such warped outer disk around SVS\,13\,A 
(see Anglada et al. 2000, 2004; the disk radius around SVS\,13\,A is estimated to be $\sim$\,32\,au).
Furthermore, in the calculations in Soker (2005), disk kinematics was apparently not taken into account. Since a warped disk should 
be rotating, it cannot simply slow down one side exclusively all the time. From the SMA CO observations, the (projected) length of 
blue bullet 2 is $\sim$\,2600\,au, and the estimated propagation time of bullet 2 is $\sim$\,310--830\,yr. Assuming that the SVS\,13\,A 
jet really interacts with the ambient gas above a warped disk at a radius of 10\,au, the period of such interaction would be 
$\sim$\,30-40\,yr (considering a Keplerian disk around a source with a mass of $\sim$\,0.8--1.0\,$M_\odot$). Therefore, during the 
propagation of blue bullet 2, the warped disk will interact roughly 10--30 times with each side of the jet. After so many times 
symmetrically interacting, the effect from the warped disk would be smoothed, and no asymmetry velocity profile would be 
expected in blue bullet 2.

Fendt (2011) suggested that helical MHD shocks can self-generate rotational motions. In this scenario, shock compression of a helical 
magnetic field results in a toroidal Lorentz force component that will accelerate the jet material in the toroidal direction. This process 
transforms magnetic angular momentum carried along the jet into kinetic angular momentum (rotation), although the jet is injected into 
the ambient gas with no initial rotation (see Fendt 2011 for details). Therefore, the motion caused by helical MHD shocks is real rotation 
and not a gradient that mimics rotation as the other alternative scenarios (i.e., twin-jets, precession, and warped disk). Nevertheless, helical 
MHD shocks only drive slow rotation, with a rotation speed of about 0.1\%--1\% of the jet bulk velocity (see Fendt 2011). The rotation 
velocity measured in blue bullet 2 is roughly 8--10\,km\,s$^{-1}$, which is about 10\% of the jet bulk velocity ($\sim$\,100\,km\,s$^{-1}$) --- 
significantly larger than that expected for rotation driven by helical MHD shocks. Hence, the probability of helical MHD shocks resulting in 
high velocity rotation of the bullets in SVS\,13\,A is extremely low.

After discussing the above alternative scenarios, the rotation of the bullets appears to be the preferred explanation left for the observed 
velocity fields in blue bullet 2 and the red bullet. However, a few uncertainties still need to be considered regarding this rotation explanation. 
The first question is why no rotation signature is seen in blue bullets 1 and 3 in the SMA CO observations. 
As introduced and discussed above, SVS\,13\,A is a FUor-like variable protostar with episodic outbursts. During outburst processes, high 
accretion rates lead to the onslaught of a large amount of material and angular momentum, which would then drive strong jets/outflows 
(due to the physical link between accretion and ejection), e.g., the bullets seen in the SVS\,13\,A jet. 
Nevertheless, theoretical studies also find large differences between episodic outbursts, with luminosity and accretion rate varying within 
at least one order of magnitude (see, e.g., Vorobyov \& Basu 2010; 2015). This is consistent with the results from numerical simulations
of protostellar outflows, in which the properties of young outflows (e.g., mass, luminosity, and momentum) vary with stellar evolution, due 
to episodic accretion (see, e.g., Machida \& Hosokawa 2013; Machida 2014).
Hence, there is a possibility that blue bullet 2 and the red bullet result from bursts with somehow more angular momentum loss (or clearer 
transverse velocity gradients), while blue bullets 1 and 3 result from relatively weaker bursts, whose underlying rotation signature is not 
easily distinguished in the observations (see below). 

As reviewed by Ray et al. (2007) and Belloche (2013), the detection of the rotation of jets/outflows is extremely difficult, not only because 
both high angular and spectral resolution are required, but also the kinematics of jets/outflows is complicated (including acceleration, rotation, 
and interaction with the environment, etc.). As shown in the velocity channel maps (see Fig.\,8), the transverse velocity gradient is not seen 
in the velocity range between $-$128 and $-$112\,km\,s$^{-1}$ for blue bullet 2 (a high-velocity front of blue bullet 2 in morphology), but much 
clearer in the velocity range between $-$110 and $-$85\,km\,s$^{-1}$ (which is shown in Figure~2). Including both velocity ranges, the velocity 
field is more complicated and the so-called `Hubble-law' effect appears in the field.
A similar situation seems to happen in blue bullet 1 (see velocity channel maps shown in Figure\,8). Although no clear transverse velocity 
gradient is detected in its high velocity emission (from $-$114 to $-$70\,km\,s$^{-1}$; dominated by the `Hubble-law' effect, see Figure~2), 
a velocity shift, away from the driving source but perpendicular to the local jet, is tentatively detected in the velocity between $-$70 and 
$-$40\,km\,s$^{-1}$. Figure~5 shows the velocity fields of blue bullet 1 in this velocity range. The direction of this tentative velocity gradient 
is consistent with that of blue bullet 2 shown in Fig.\,2, i.e., the left side of the local jet axis is blueshifted while the right side is redshifted. 
So, there is a possibility that blue bullet 1 is also rotating, but its rotation signature (i.e., transverse velocity gradient) is more hidden or 
confused by its complicated kinematics (compared with blue bullet 2), which can only be indistinctly seen in a certain velocity range.
We also note that the comparison between the SMA CO observations (this work) and the Keck [Fe~II] observations (Hodapp \& Chini 2014) 
shows a tentative velocity gradient across the microjet detected in the Keck images, which is consistent with that found in blue bullet 2 in 
the SMA CO images (see Appendix~B). 

Another uncertainty of the rotation explanation comes from the driving disk of the jet. To our knowledge, the kinematics of the circumstellar 
disk around SVS\,13\,A is unknown yet. As suggested by Belloche (2013), one important requirement for convincingly detecting rotation in 
jets is that the velocity gradients should be consistent between both outflow lobes and with the disk rotation. Due to insufficient observations, 
this requirement is still missing in the SVS\,13\,A jet system. 

In summary, due to the uncertainties discussed above, the detection of the rotation of the bullets in the SVS\,13\,A jet is {\it not} definitive yet. 
Based on the SMA CO\,(2--1) observations, we consider the rotation of the bullets thus far the most likely explanation for the observed velocity 
fields in blue bullet 2 and the red bullet. Hereafter we analyze the properties of the two bullets according to this explanation. It must be noted 
that further high angular resolution observations are needed to comprehensively study the kinematics of the SVS\,13\,A jet, as well as the disk 
of the driving source, in order to confirm the rotation nature of these bullets.

\subsection{Angular Momenta in the Bullets}

For the EHV bullets in the SVS\,13\,A jet (velocities $>$ 70\,km\,s$^{-1}$ and no $^{13}$CO line emission detected at these velocity 
ranges), the CO\,(2--1) emission can be safely regarded as optically thin. The gas masses of the bullets are then derived with the same 
method as described in Cabrit \& Bertout (1990). In the calculations, we assume LTE conditions, and excitation temperatures of 25\,K 
and 45\,K for the blue and red bullets, respectively, which were derived from single-dish CO\,(2--1)/(1--0) observations (see Masson 
et al. 1990). The estimated bullet masses are then $\sim$\,3.8\,$\times$\,10$^{-4}$ and $\sim$\,3.1\,$\times$\,10$^{-4}$\,$M_\odot$ for 
blue bullet 2 and the red bullet, respectively. The masses of the other bullets, as well as other jet properties, will be presented and discussed 
in future work. 

Based on the SMA CO PV diagrams (see Figure~3), rotation velocity $v_{\phi}$ and radius $R_{\rm rot}$ of the two bullets are measured 
between the blue and red peaks in the PV diagrams, corrected for the inclination angle ($\it i$ = 40$^\circ$), i.e.,
\begin{equation}
v_{\rm \phi} = \frac{1}{\sin i} \frac{v_{\rm shift}}{2},
\end{equation}
\begin{equation}
R_{\rm rot} = \frac{L_{\rm shift}}{2}.
\end{equation}
\noindent 
The measured rotation velocities and radii are $\sim$\,11.7--13.7\,km\,s$^{-1}$ and $\sim$\,440--610\,au for blue bullet 2, and
$\sim$\,4.7\,$\pm$\,0.5\,km\,s$^{-1}$ and $\sim$\,1200\,$\pm$\,200\,au for the red bullet. These rotation velocities are comparable
to the values measured in the rotating jet in DG~Tau (6--15\,km\,s$^{-1}$; Bacciotti et al. 2002), but much larger than those measured 
in other rotating outflow candidates, e.g., CB\,26 (1--2\,km\,s$^{-1}$; Launhardt et al. 2009). 
The estimated local specific angular momentum ($v_{\phi}$\,$\times$\,$R_{\rm rot}$) for the two rotating bullets is 
$j_{\rm blue~bullet}$\,$\sim$\,9.8\,$\times$\,10$^{21}$\,cm$^{2}$\,s$^{-1}$ and 
$j_{\rm red~bullet}$\,$\sim$\,8.5\,$\times$\,10$^{21}$\,cm$^{2}$\,s$^{-1}$, 
respectively. The specific angular momenta of the two bullets are much higher than that of the rotating jet in DG~Tau 
($\sim$\,3.5\,$\times$\,10$^{20}$\,cm$^{2}$\,s$^{-1}$; Bacciotti et al. 2002) and that of the rotating outflow in CB\,26 
($\sim$\,1.5\,$\times$\,10$^{20}$\,cm$^{2}$\,s$^{-1}$; Launhardt et al. 2009). It may imply that rotating jets from episodic outbursts in 
the protostellar phase can more efficiently remove the (specific) angular momentum from the star-disk system, compared with outflows 
in the T~Tauri phase.  

For a further comparison, we present the two rotating bullets in Figure~6, which shows the distribution of specific angular momentum as a 
function of rotation radius for different scales of a star forming region -- from molecular clouds, through protostellar envelopes, circumstellar 
disks, and binaries, all the way to the Sun. This figure is mainly based on the diagram published by Belloche (2013, and references therein). 
As seen in Figure~6, the specific angular momenta of the two rotating bullets are comparable to those of dense cores, about 10 times larger
than those of protostellar envelopes, and about 20 times larger than those of circumstellar disks (also corrected for inclination; see Belloche 
2013). The large amount of specific angular momenta associated with the two bullets indicates that the rotation of the jets/outflows may be
the most efficient method in extracting angular momentum in the early phases of star formation.
It should be noted that the total angular momenta (mass being taken into account) of these circumstellar disks are still larger than those of 
the bullets, considering that the disk mass of embedded protostars (disk masses ranging from $\sim$\,0.02 to $\sim$\,0.1\,$M_\odot$ with 
a median of 0.04\,$M_\odot$; see a review by Williams \& Cieza 2011) is hundreds of times larger than the masses of the bullets. Note that, 
for the SVS\,13\,A jet, the disk mass of the driving source was estimated to be $\sim$\,0.06\,$M_\odot$ (Anglada et al. 2004).

The comparison suggests that most of the angular momentum of a protostellar circumstellar disk can be extracted by only a few tens of such 
rotating bullets produced by episodic outbursts. Interestingly, based on the IRAS survey, Kenyon et al. (1990) estimated that each YSO 
experiences at least 10 FUor-type outbursts during its evolution. Recent theoretical simulations also suggest that YSOs will experience roughly 
20--30 outbursts during the early phase of star formation (see, e.g., Vorobyov \& Basu 2015). The  tentative discovery of rotating bullets in 
the SVS\,13\,A jet, as well as their large angular momentum, would be consistent with this episodic accretion/burst scenario in star formation.

\subsection{Launching Mechanism of Bullets}

The precise origin of jets from YSOs is still hotly debated. Theoretical studies have suggested several launching models for jets, 
including stellar wind, X-wind, and disk-wind (see Pudritz et al. 2007; Shang et al. 2007; Frank et al. 2014). The jet launching radii,
depending on the launching models, range from a few $R_\odot$ (stellar wind), $\leq$\,0.1\,au (X-wind), to possibly several au 
(disk-wind). As reviewed by Ray et al. (2007) and Frank et al. (2014), a key observational diagnostic to discriminate between these 
models is to search for signatures of rotation in jets/outflows, and then to infer the launching `footprint'. Thus far most studies, which 
interpret the observed velocity gradients perpendicular to the outflow axis as due to jet rotation, seem to be consistent with an origin 
from an extended disk-wind, launched from between 0.1 to 3--5\,au (see Cabrit 2009; Frank et al. 2014). However, it must be noted 
that these observations cannot exclude the existence of inner stellar or X-winds because only external structures of the jet have been 
observed due to limited angular resolution.

We also estimate the launching footprint of the SVS\,13\,A jet based on the presumed rotational signature of blue bullet 2 (which has 
a more systematic velocity field and higher signal-to-noise ratio, as compared with the red bullet), using the equation suggested by 
Anderson et al. (2003): 
\begin{equation}
R_{0} \approx 0.7\,{\rm au} \left(\frac{R}{10~\rm au}\right)^{2/3} \left(\frac{v_{\phi, R}}{10~\rm km\,s^{-1}}\right)^{2/3} \left(\frac{v_{p, R}}{100~\rm km\,s^{-1}}\right)^{-4/3} \left(\frac{M_{\star}}{1~M_\odot\,}\right)^{1/3}.
\end{equation}
\noindent In the equation, $v_{\phi, R}$ and $v_{p, R}$ are toroidal and poloidal velocities at a distance $R$ from the jet axis, respectively. 
Based on the SMA CO PV diagrams (see Figure~3), toroidal and poloidal velocities are measured using the following equations, 
\begin{equation}
v_{\phi, R} =  \frac{1}{\sin i} \frac{v_{\rm blue~peak} - v_{\rm red~peak}}{2},
\end{equation}
\begin{equation}
v_{p, R} =  \frac{1}{\cos i} \frac{v_{\rm blue~peak} + v_{\rm red~peak}}{2}.
\end{equation}
\noindent In the calculations, the mass of the driving source $M_{\star}$ is adopted as $\sim$\,1.0\,$M_\odot$ ( see Chen et al. 2009) 
and inclination angle $i$ as 40$^\circ$. The estimated launching footprint for blue bullet 2 is then $\sim$\,7.2--7.7\,au, which appears 
to support the extended disk-wind model. Note that the disk radius of SVS\,13\,A was estimated to $\sim$\,32\,au (Anglada et al. 2004), 
which is about 4--5 times larger than the launching radius for blue bullet 2. 

\section{SUMMARY}

We present SMA CO\,(2--1) observations toward the well-known prototellar jet driven by SVS\,13\,A, a variable protostar in the NGC\,1333 
star-forming region. SMA CO images show high-velocity jets from SVS\,13\,A, with velocities ranging from $\sim$\,$-$150 to $\sim$\,150\,km\,s$^{-1}$. 
In this work, we focus on the extremely high velocity (EHV) components (jet velocities $>$ 70\,km\,s$^{-1}$ over the cloud systemic velocity), 
the so-called molecular bullets. The main results of this work are summarized below.

(1) Three bullets are found in the blueshifted jet (blue bullets), and one bullet is found in the redshifted jet (red bullet). The blueshifted jet 
shows a wiggle morphology, which can be reasonably fitted by an orbiting-jet model. The best-fit parameters suggest that the source itself 
is a binary system with an orbital radius of 0\farcs16\,$\pm$\,0\farcs02, which is consistent with previous VLA observations. Considering that 
SVS\,13\,A is a FUor-like variable protostar, we suggest that the bullets in the SVS\,13 jet are formed by periodic bursts of mass, which take 
place every few hundred years.

(2) Based on the SMA CO\,(2--1) observations, we discover in one blueshifted bullet (blue bullet 2) and the redshifted bullet clear and large 
systematic velocity gradients, perpendicular to the jet axis. The velocity gradients in the blue- and redshifted bullets are roughly in the same 
direction. After discussing several scenarios, such as twin-jets, jet precession, warped disk, and helical shock, we suggest that the observed 
velocity gradients in the bullets most likely result from the rotation of the SVS\,13\,A jet.

(3) From the SMA CO images, the measured rotation velocities are $\sim$\,11.7--13.7\,km\,s$^{-1}$ for the blueshifted bullet and 
$\sim$\,4.7\,$\pm$\,0.5\,km\,s$^{-1}$ for the redshifted bullet. The estimated specific angular momenta are 
$\sim$\,9.8\,$\times$\,10$^{21}$\,cm$^{2}$\,s$^{-1}$ for the blueshifted bullet and $\sim$\,8.5\,$\times$\,10$^{21}$\,cm$^{2}$\,s$^{-1}$ 
for the redshifted bullet, which are comparable with those of dense cores, about 10 times larger than those of protostellar envelopes and 
about 20 times larger than those of circumstellar disks. The significant amount of specific angular momenta associated with the 
two bullets indicates that the rotation of the jets/outflows is an efficient method in extracting angular momentum in the early phases 
of star formation, which may be a key mechanism to resolve the ``angular momentum problem" in the field of star formation.

(4) Based on the jet kinematics derived from SMA CO observations, the estimated launching footprint on the disk has a radius of 
$\sim$\,7.2--7.7\,au, which appears to support the extended disk-wind model. 

(5) Nevertheless, due to remaining uncertainties (e.g., absence of clear signs of rotation in the other bullets), the detection of the jet 
rotation in SVS\,13\,A is not definitive yet. Further observations are needed to comprehensively understand the kinematics of the 
SVS\,13\,A jet system (including circumstellar disk), in order to confirm the rotation nature of the bullets.

\acknowledgments

We thank the anonymous referee for providing many insightful suggestions and comments, which helped us to improve this work.
We acknowledge the SMA staff for technical support during the observations. We thank K. W. Hodapp for providing us Keck [FeII] 
data. This work was supported by the National Natural Science Foundation of China (grants Nos. 11473069 and 11328301) and 
the Strategic Priority Research Program of the Chinese Academy of Sciences (grant No. XDB09000000). X.C. acknowledges the 
support of the Thousand Young Talents Program of China.


\clearpage

\appendix

\section{SMA CO\,(2--1) Velocity Channel Maps of Bullets}

In this Appendix, we present the SMA CO\,(2--1) velocity channel maps of blue bullet 3 (Figure~7), blue bullets 2 \& 1 (Figure~8), and
the red bullet (Figure~9). For all these channel maps, the phase center is R.A. = 03$^{\rm h}$29$^{\rm m}$03$^{\rm s}$.07, 
decl. = 31$^\circ$15$'$52$\farcs$00 (J2000). The center velocity of the channel is written in the top left corner of each panel (in 
km\,s$^{-1}$), while the filled grey ellipse (lower right corner) in each panel indicates the SMA synthesized beam. The systemic velocity 
of the cloud is $\sim$\,9\,km\,s$^{-1}$. 

\section{The Comparison Between SMA CO and Keck [Fe~II] Data}

In this Appendix, we compare the SMA CO\,(2--1) observations with the Keck [Fe~II] 1.644\,$\mu$m near-infrared observations. The inner 
microjet in SVS\,13\,A was observed in the [Fe~II] line at the Keck telescopes in 2012 and 2013 (Hodapp \& Chini 2014). Interestingly, the 
2012 Keck image shows a tentative detection of a velocity gradient across the driving source (see Figure~10). This velocity gradient is 
perpendicular to the microjet and is in the same direction as those in blue bullet 2 and red bullet seen in the SMA CO\,(2--1) observations. In 
this context, the velocity gradient seen in the 2012 Keck image could be interpreted as the rotation of the microjet, in comparison with the 
SMA CO velocity fields of the bullets. However, it must be noted that the velocity gradient seen in the 2012 Keck image is not seen in the 
2013 Keck image. Figure~11 shows a comparison between the 2012 and 2013 Keck data. As seen in Fig.\,11, the 2013 data around the 
driving source do not extend as much as the 2012 data (top panels); furthermore, the noise level around the driving source is about 
10--20\,km\,s$^{-1}$ in the 2012 image, but about 30\,km\,s$^{-1}$ in the 2013 image (bottom panels). The velocity shift across the driving 
source seen in the 2012 image is about 60-70\,km\,s$^{-1}$, which is only $\sim$\,2 times larger than the noise levels in the 2013 image. 
Therefore, the absence of the velocity gradient in the 2013 image is probably due to: (1) the region that appears to show a velocity gradient 
is not fully detected in the 2013 observations, and (2) relatively lower signal-to-noise ratios in this region. 

We note that Hodapp \& Chini (2014) tried to find rotation signature in the SVS\,13\,A microjet, but concluded they do not detect any rotation. 
Indeed, including both the 2012 and 2013 Keck data, it will be difficult to distinguish any velocity gradient, due to the difference between the 
two epoch data.  Further high angular resolution observations at the Keck telescopes (or other large telescopes) are needed to confirm the 
detection of the velocity gradient across the driving source seen in the 2012 Keck data. 

\clearpage

\clearpage


\begin{figure*}
\begin{center}
\figurenum{1}
\includegraphics[width=14cm,angle=0]{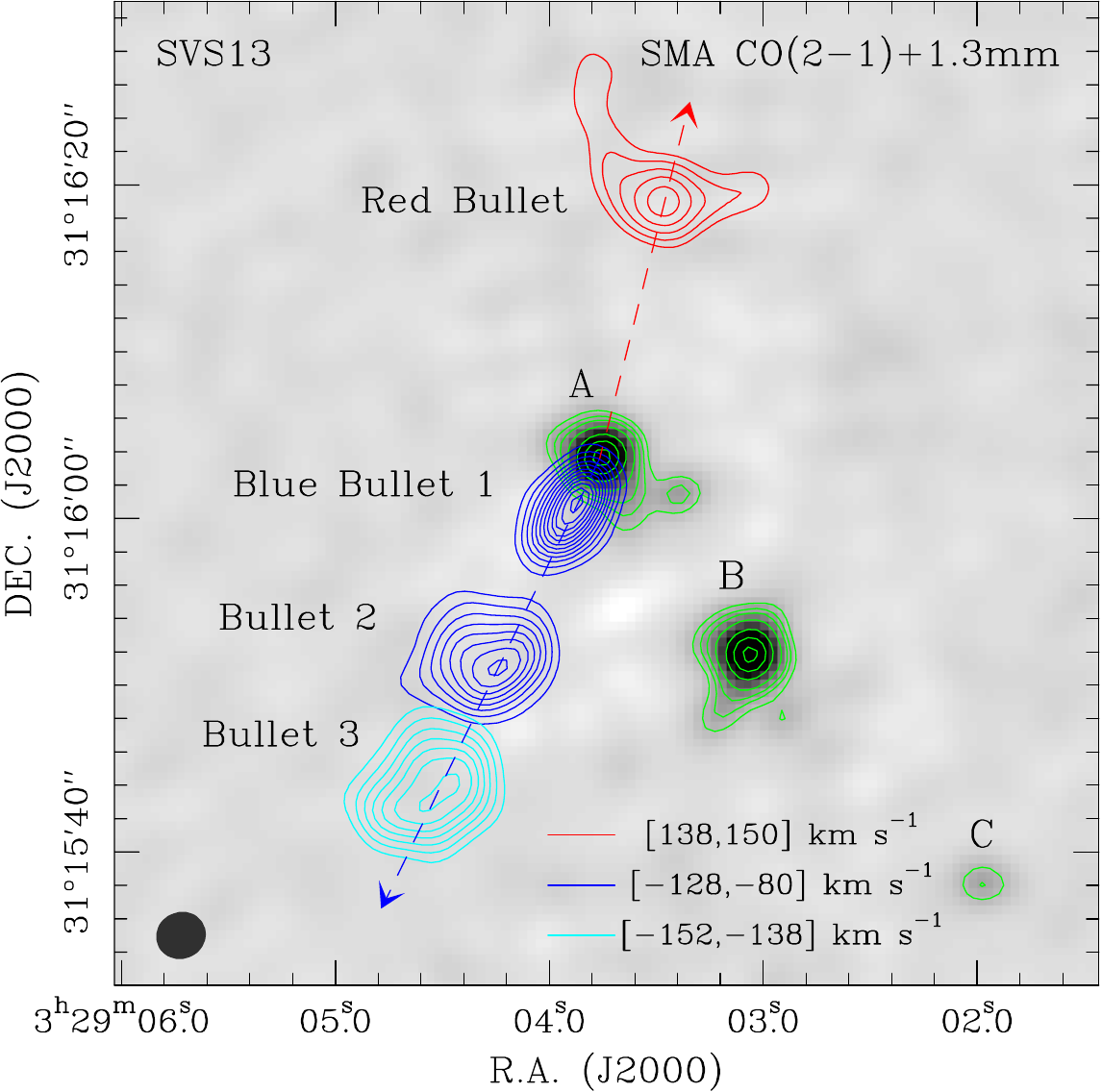}
\caption{Velocity-integrated intensity maps of the SMA CO\,(2--1) emission from SVS\,13, plotted on the SMA 1.3\,mm dust continuum 
image (grey image and green contours; from Chen et al. 2013). The red, blue, and cyan contours represent CO\,(2--1) emission 
integrated over three different velocity ranges shown in the image, which is redshifted or blueshifted with respect to the cloud systemic 
velocity ($\sim$\,9\,km\,s$^{-1}$). The CO red contours start at 0.64\,Jy\,beam$^{-1}$\,km\,s$^{-1}$ ($\sim$\,4\,$\sigma$) and increase 
in steps of 0.32\,Jy\,beam$^{-1}$\,km\,s$^{-1}$ ($\sim$\,2\,$\sigma$), blue contours start at 1.6\,Jy\,beam$^{-1}$\,km\,s$^{-1}$ 
($\sim$\,4\,$\sigma$) and increase in steps of 1.2\,Jy\,beam$^{-1}$\,km\,s$^{-1}$ ($\sim$\,3\,$\sigma$), while cyan contours start at 
0.64\,Jy\,beam$^{-1}$\,km\,s$^{-1}$ ($\sim$\,4\,$\sigma$) and increase in steps of 0.48\,Jy\,beam$^{-1}$\,km\,s$^{-1}$ ($\sim$\,3\,$\sigma$).
The three dust continuum sources found in the region are named A, B, and C, respectively (see Chen et al. 2013).
The blue and red dashed line arrows show the overall directions of the blueshifted and redshifted jets from source SVS\,13\,A. The grey 
oval in the bottom left corner indicates the SMA synthesized beam.
\label{outflow}}
\end{center}
\end{figure*}

\begin{figure*}
\begin{center}
\figurenum{2}
\includegraphics[width=10cm,angle=0]{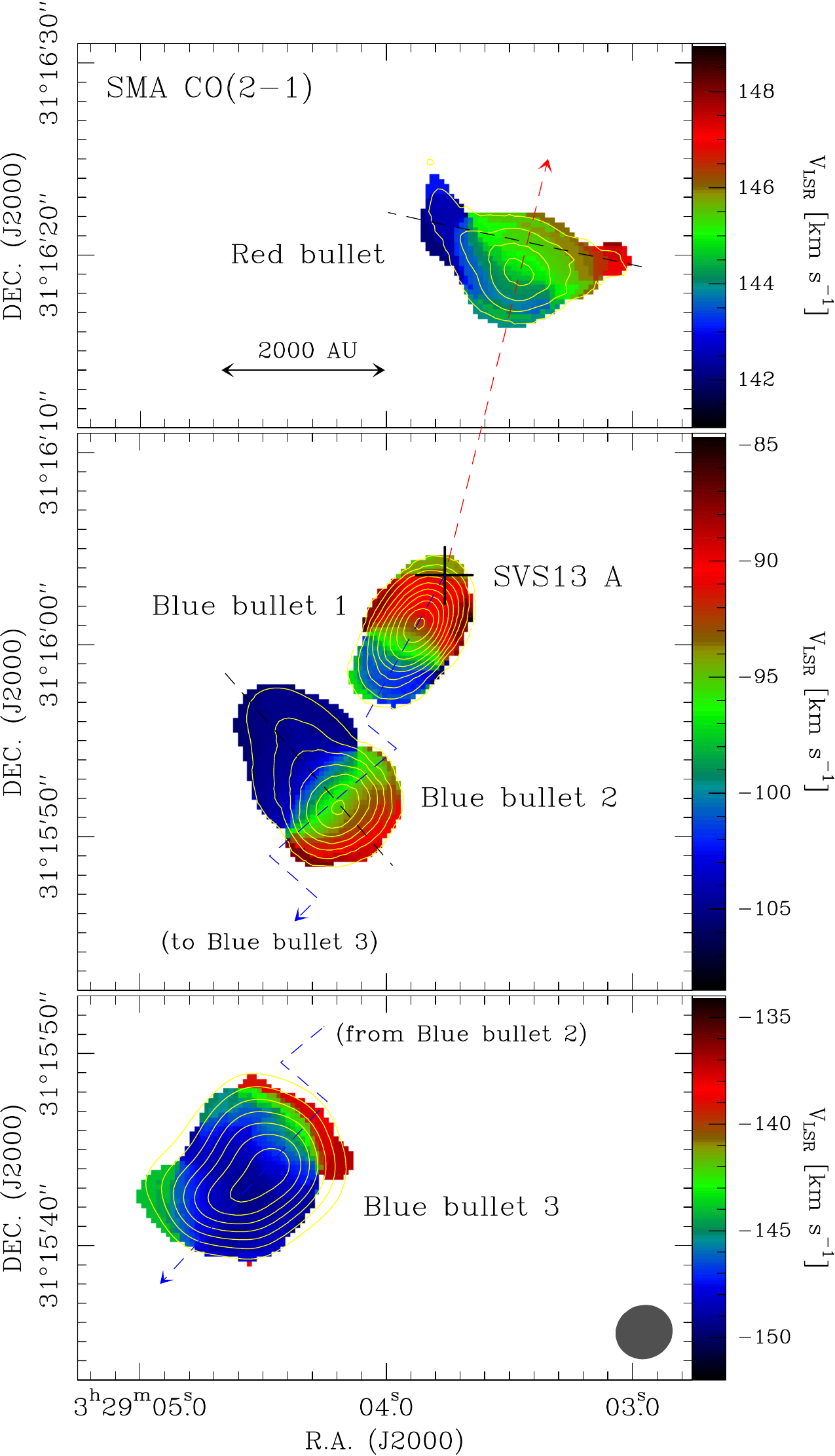}
\caption{\footnotesize SMA CO\,(2--1) integrated intensity map (yellow contours) and mean velocity field (1st moment velocity map, color shades) 
of the SVS\,13\,A jet. {\it Top panel:} Red bullet (velocities integrated from +142 to +150\,km\,s$^{-1}$). Contours start at 4\,$\sigma$ and then increase 
in steps of 3\,$\sigma$ (1\,$\sigma$\,$\sim$\,0.16\,Jy\,beam$^{-1}$\,km\,s$^{-1}$). {\it Middle panel:} Blue bullets 1 and 2 (velocities integrated from 
$-$106 to $-$83\,km\,s$^{-1}$). Contours start at 4\,$\sigma$ and then increase in steps of 4\,$\sigma$ (1\,$\sigma$\,$\sim$\,0.30\,Jy\,beam$^{-1}$\,km\,s$^{-1}$).
{\it Bottom panel:} Blue bullet 3 (velocities integrated from $-$154 to $-$134\,km\,s$^{-1}$). Contours start at 3\,$\sigma$ and then increase in steps of 
3\,$\sigma$ (1\,$\sigma$\,$\sim$\,0.18\,Jy\,beam$^{-1}$\,km\,s$^{-1}$). The blue dashed line arrows shows the jet axis defined by the jet wiggle (see text). 
The cross marks the peak position of the dust continuum source, while the grey oval in the bottom right corner indicates the SMA synthesized beam.
\label{velocity}}
\end{center}
\end{figure*}

\begin{figure*}
\begin{center}
\figurenum{3}
\includegraphics[width=10cm,angle=0]{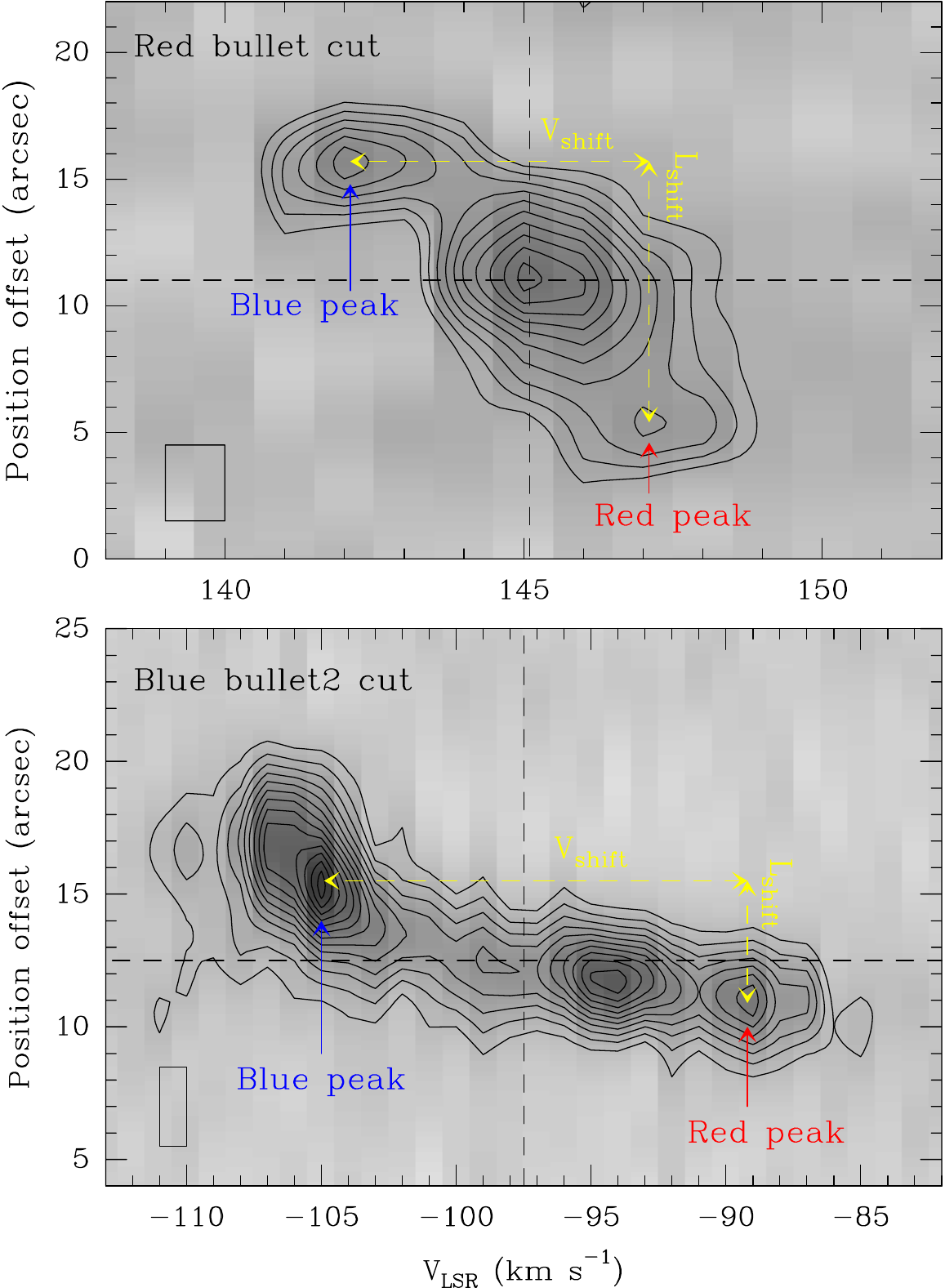}
\caption{Position-velocity diagrams of the transversal cuts marked in Figure~2. {\it Top panel:} Cut across the red bullet.
Contours start at 84\,mJy\,beam$^{-1}$ ($\sim$\,3\,$\sigma$) and increase in steps of 28\,mJy\,beam$^{-1}$ ($\sim$\,1\,$\sigma$).
{\it Bottom panel:} Cut across the blue bullet 2. Contours start at 108\,mJy\,beam$^{-1}$ ($\sim$\,3\,$\sigma$) and increase in 
steps of 72\,mJy\,beam$^{-1}$ ($\sim$\,2\,$\sigma$). The yellow dashed line arrows show the velocity shift (V$_{\rm shift}$) and 
shift distance (L$_{\rm shift}$) between the blue and red peaks. The rectangle in the bottom left of the two panels shows the velocity 
and angular resolutions of the observations. The two black dashed lines mark the position of the symmetry axis and radial mean velocity 
of the two bullets.
\label{pv_bullets}}
\end{center}
\end{figure*}

\begin{figure*}
\begin{center}
\figurenum{4}
\includegraphics[width=11cm,angle=0]{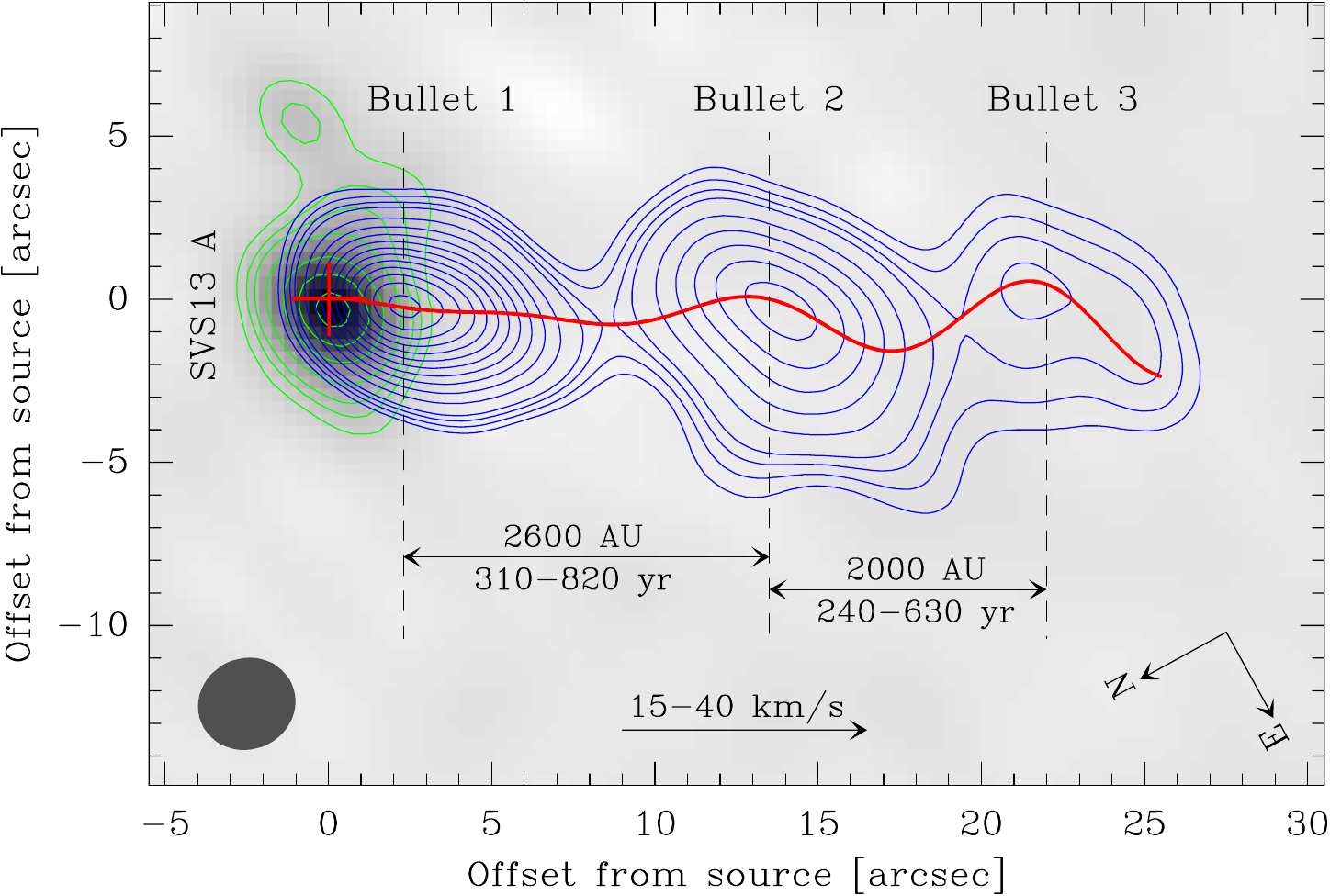}
\caption{Velocity-integrated intensity map of the SMA CO\,(2--1) emission from SVS\,13\,A (blue contours), plotted on the 
SMA 1.3\,mm dust continuum image (grey image and green contours). The maps have been rotated by 241.2 degree 
clockwise. The CO emission is integrated over the velocities ranging from $-$152 to $-$70\,km\,s$^{-1}$, and contours 
correspond to 8, 12, 16, 20\,$\sigma$ and then increase in steps of 10\,$\sigma$ (1\,$\sigma$\,$\sim$\,0.16\,Jy\,beam$^{-1}$\,km\,s$^{-1}$). 
The red cross marks the peak position of the SVS\,13\,A dust continuum source, while the red wavy curve shows the best-fit of the 
orbiting source jet model to the CO integrated emission. The black arrow in the bottom shows the proper motions of the jet 
(in a velocity range of 15--40\,km\,s$^{-1}$; see text), while the periods shown in the map are estimated from the spacings between
the bullets and the proper motions of the jet.
\label{wiggle}}
\end{center}
\end{figure*}

\begin{figure*}
\begin{center}
\figurenum{5}
\includegraphics[width=15cm,angle=0]{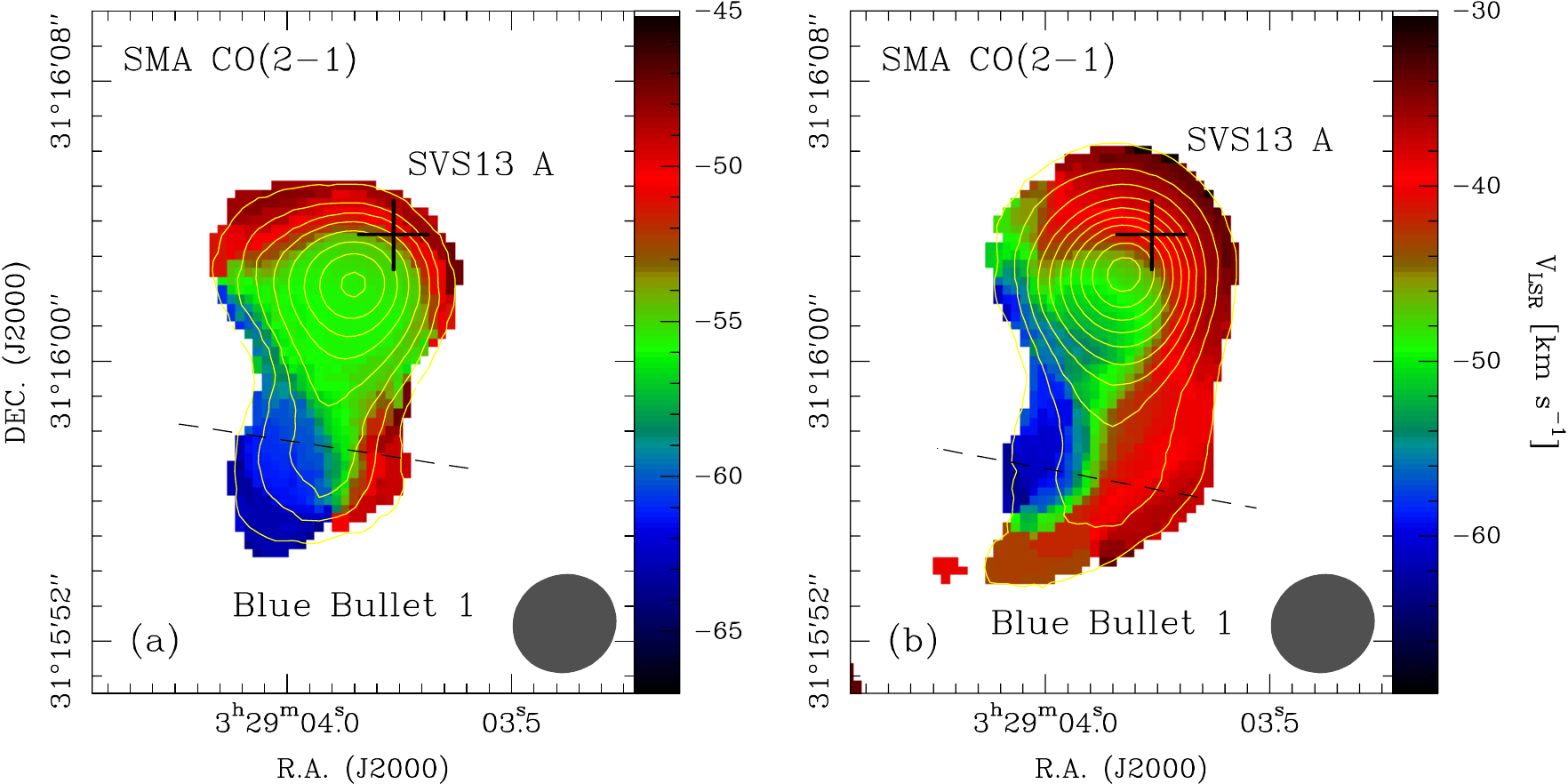}
\caption{(a) SMA CO\,(2--1) velocity-integrated intensity map (yellow contours) and mean velocity field (1st moment velocity map, color shades) 
of the SVS\,13\,A jet (velocities from $-$66 to $-$47\,km\,s$^{-1}$). Contours start at 5\,$\sigma$ and then increase in steps of 4\,$\sigma$ 
(1\,$\sigma$\,$\sim$\,0.24\,Jy\,beam$^{-1}$\,km\,s$^{-1}$). (b) Similar as panel a, but velocities integrated from $-$75 to $-$30\,km\,s$^{-1}$. 
Contours start at 6\,$\sigma$ and then increase in steps of 5\,$\sigma$ (1\,$\sigma$\,$\sim$\,0.43\,Jy\,beam$^{-1}$\,km\,s$^{-1}$). In the two panels, 
the cross marks the peak position of the dust continuum source SVS\,13\,A, while the grey oval in the bottom right corner indicates the SMA  
synthesized beam. The dashed lines show the directions of the tentative velocity gradients seen across the low-velocity part of blue bullet 1.
\label{newvelocity}}
\end{center}
\end{figure*}

\begin{figure*}
\begin{center}
\figurenum{6}
\includegraphics[width=12cm,angle=0]{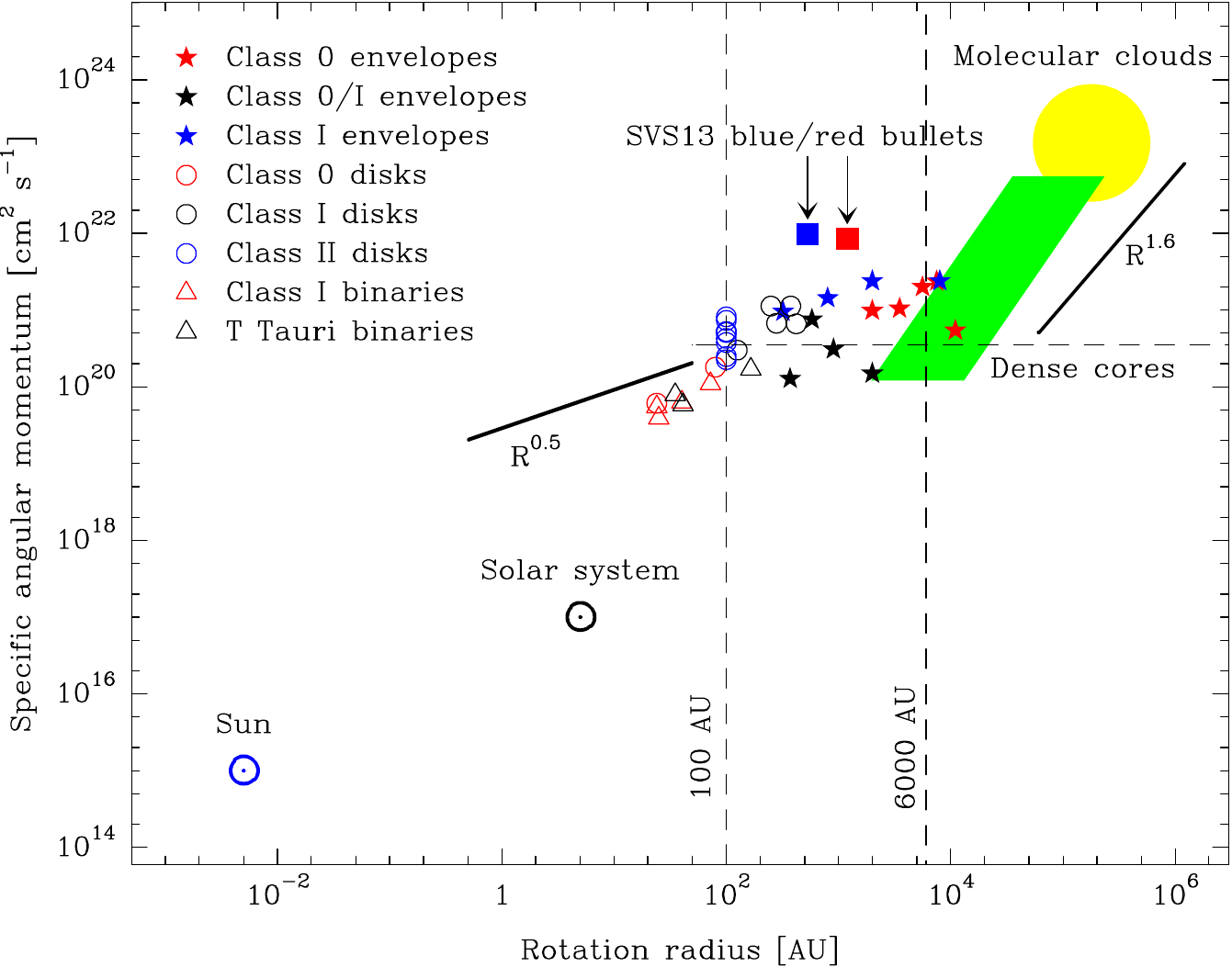}
\caption{Specific angular momentum as a function of different scales of a star forming region. The specific angular momentum
is the local value (i.e., the product of the rotation velocity and the radius) for all categories except for the binaries, the solar system, 
and the Sun for which it corresponds to the mean angular momentum per unit of mass (see Belloche 2013 and references therein). 
Yellow filled circle and green filled parallelogram show the approximate positions for the distributions of molecular clouds and dense
cores, respectively. For all disks and most of protostellar envelopes, as well as the two rotating bullets from SVS\,13\,A (red and blue 
squares), the values have been corrected for inclination. The two solid lines are shown to guide the eye; they are not least-square 
fits to the data. The vertical dashed lines mark the approximate position of breaks in the distribution of angular momentum and the 
horizontal one indicates the typical specific angular momentum during the protostellar collapse phase (see Belloche 2013 for more 
details).\label{angular_momentum}}
\end{center}
\end{figure*}

\begin{figure*}
\begin{center}
\figurenum{7}
\includegraphics[width=16cm,angle=0]{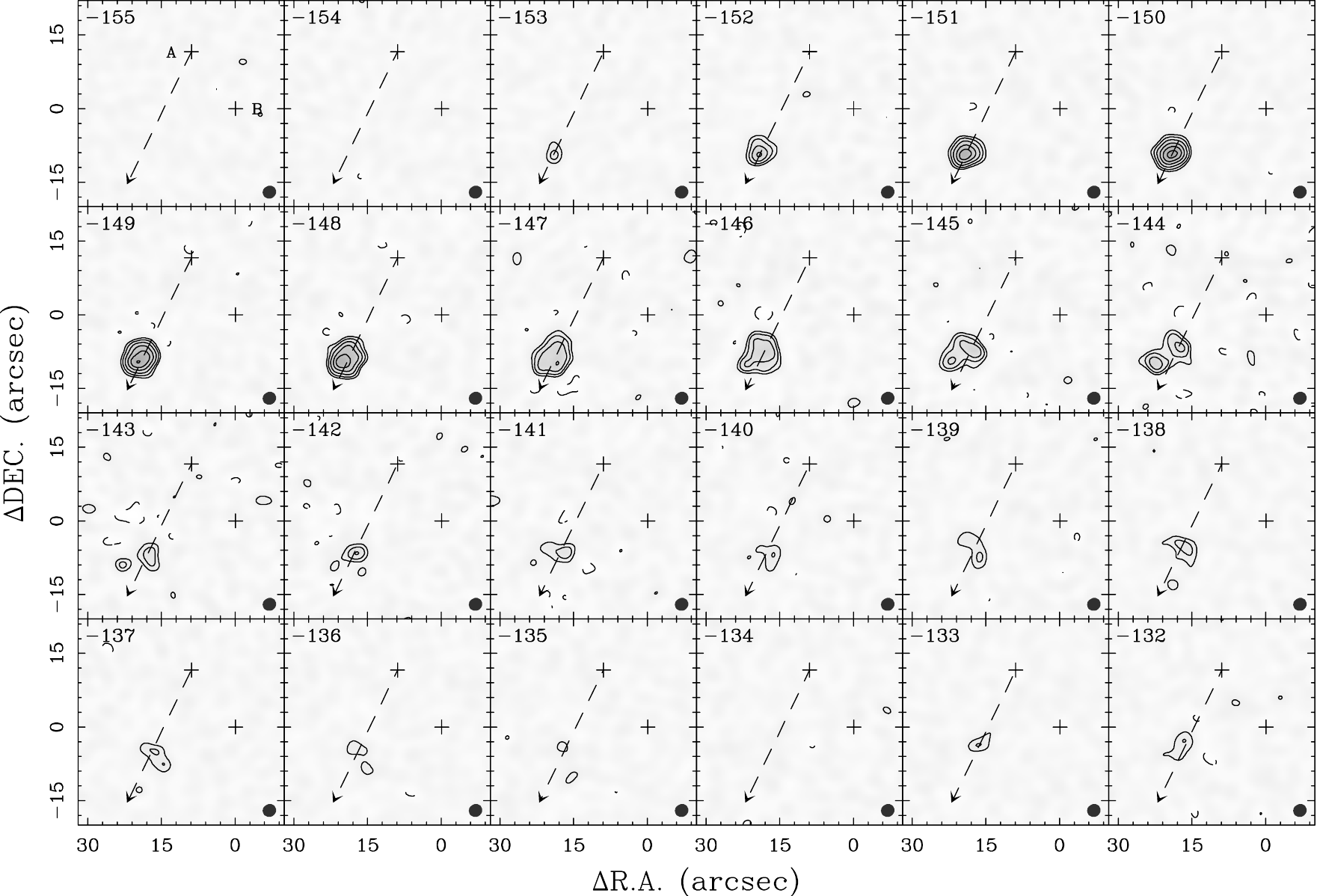}
\caption{SMA CO\,(2--1) velocity channel map of blue bullet~3. Contours levels correspond to -3, 3, 5, 8, 12, 16, and 20\,$\sigma$, then 
increase in steps of 5\,$\sigma$, where the 1\,$\sigma$ level is $\sim$\,32\,mJy beam$^{-1}$. In each panel, the two crosses mark the 
peak positions of the dust continuum sources (SVS\,13\,A and B). The dashed line arrow in each panel shows the overall direction of the 
SVS\,13\,A blueshifted jet.\label{channel_map_bullet3}}
\end{center}
\end{figure*}

\begin{figure*}
\begin{center}
\figurenum{8a}
\includegraphics[width=16cm,angle=0]{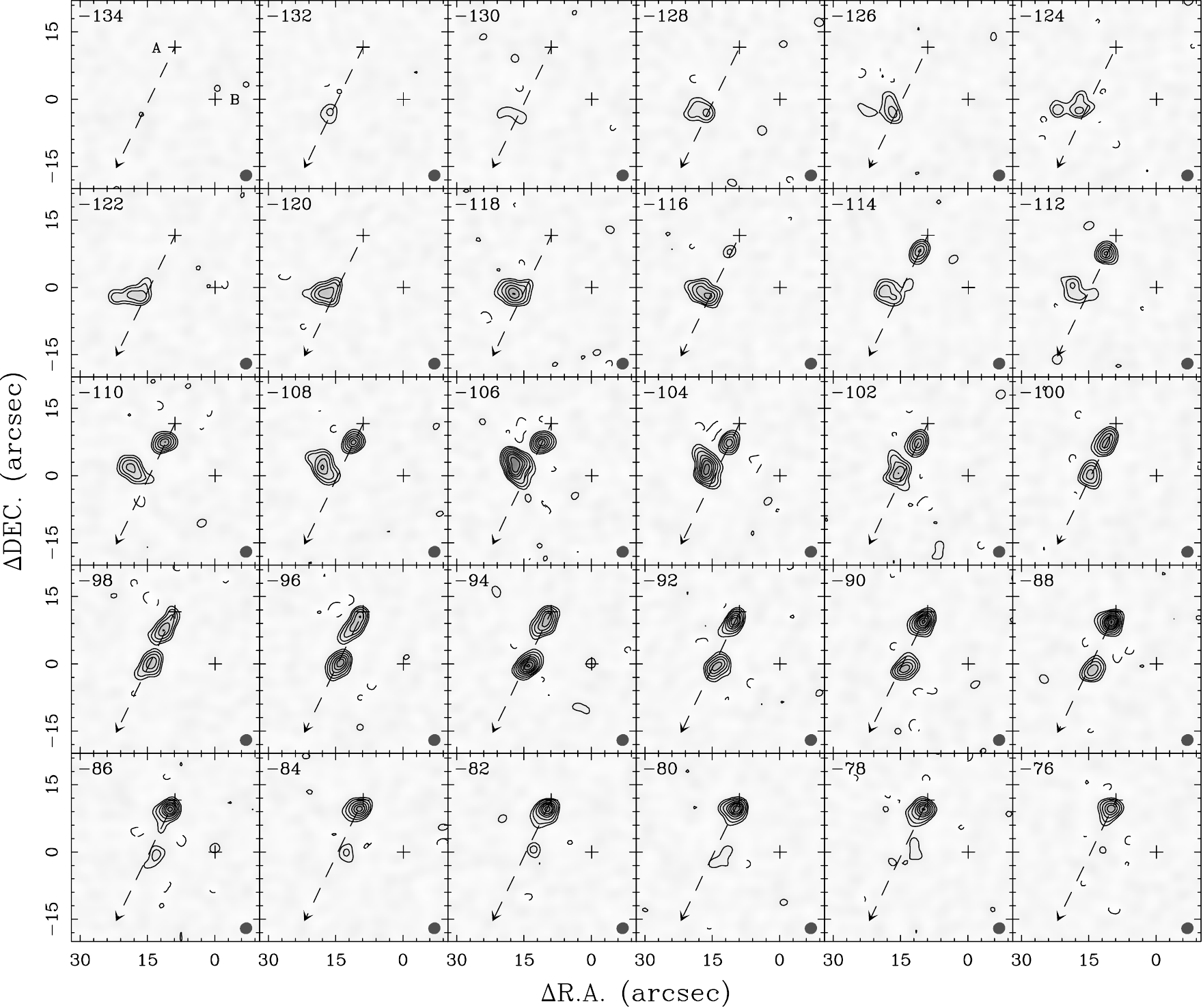}
\caption{SMA CO\,(2--1) velocity channel map of blue bullets~2 \& 1. Contours levels correspond to -3, 3, 5, 8, 12, 16, and 20\,$\sigma$, 
then increase in steps of 5\,$\sigma$, where the 1\,$\sigma$ level is $\sim$\,25\,mJy beam$^{-1}$. In each panel, the two crosses mark the 
peak positions of the dust continuum sources (SVS\,13\,A and B). The dashed line arrow in each panel shows the overall direction of the 
SVS\,13\,A blueshifted jet.\label{channel_map_bullet2}}
\end{center}
\end{figure*}

\begin{figure*}
\begin{center}
\figurenum{8b}
\includegraphics[width=16cm,angle=0]{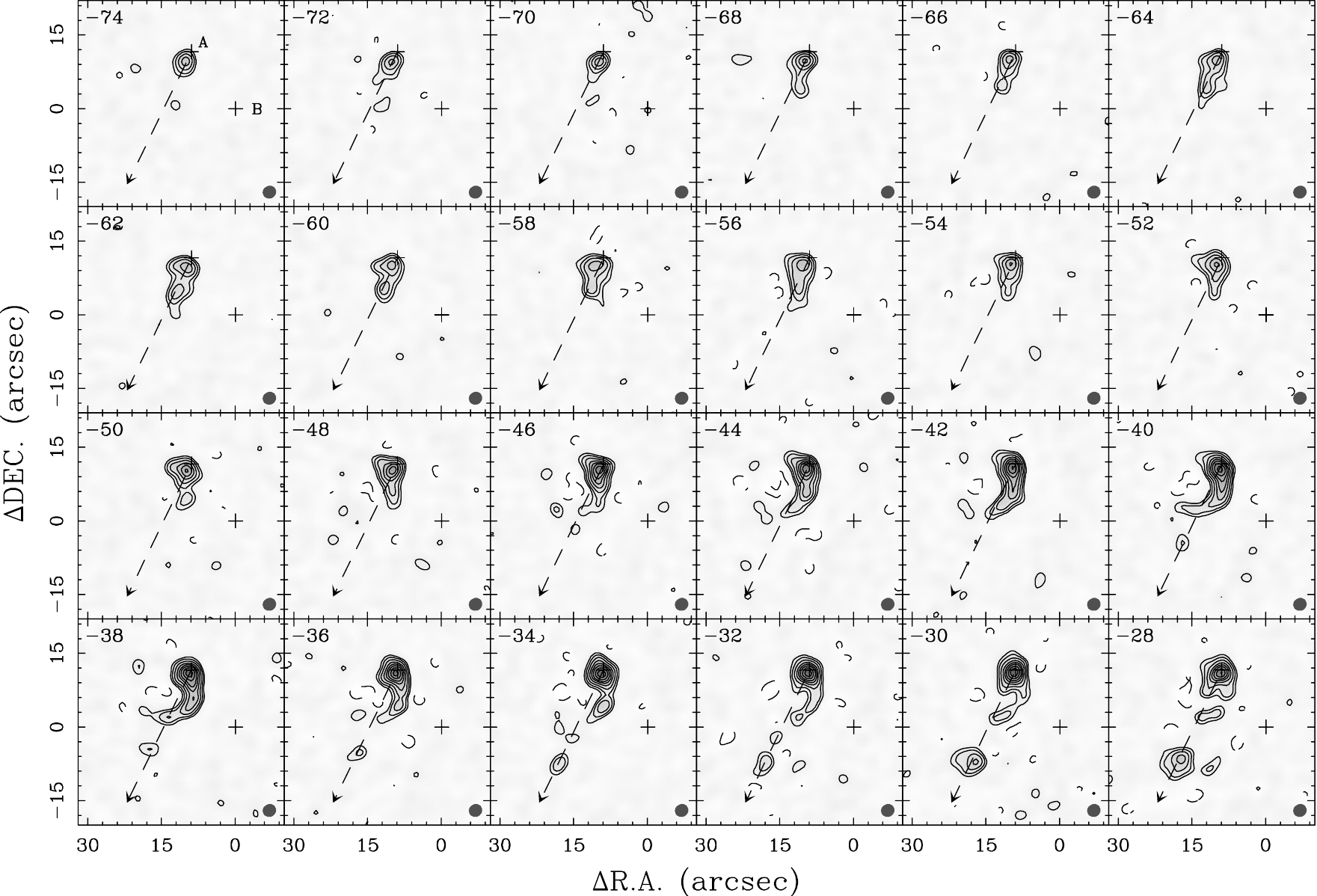}
\caption{(continued).\label{channel_map_bullet1}}
\end{center}
\end{figure*}

\begin{figure*}
\begin{center}
\figurenum{9}
\includegraphics[width=16cm,angle=0]{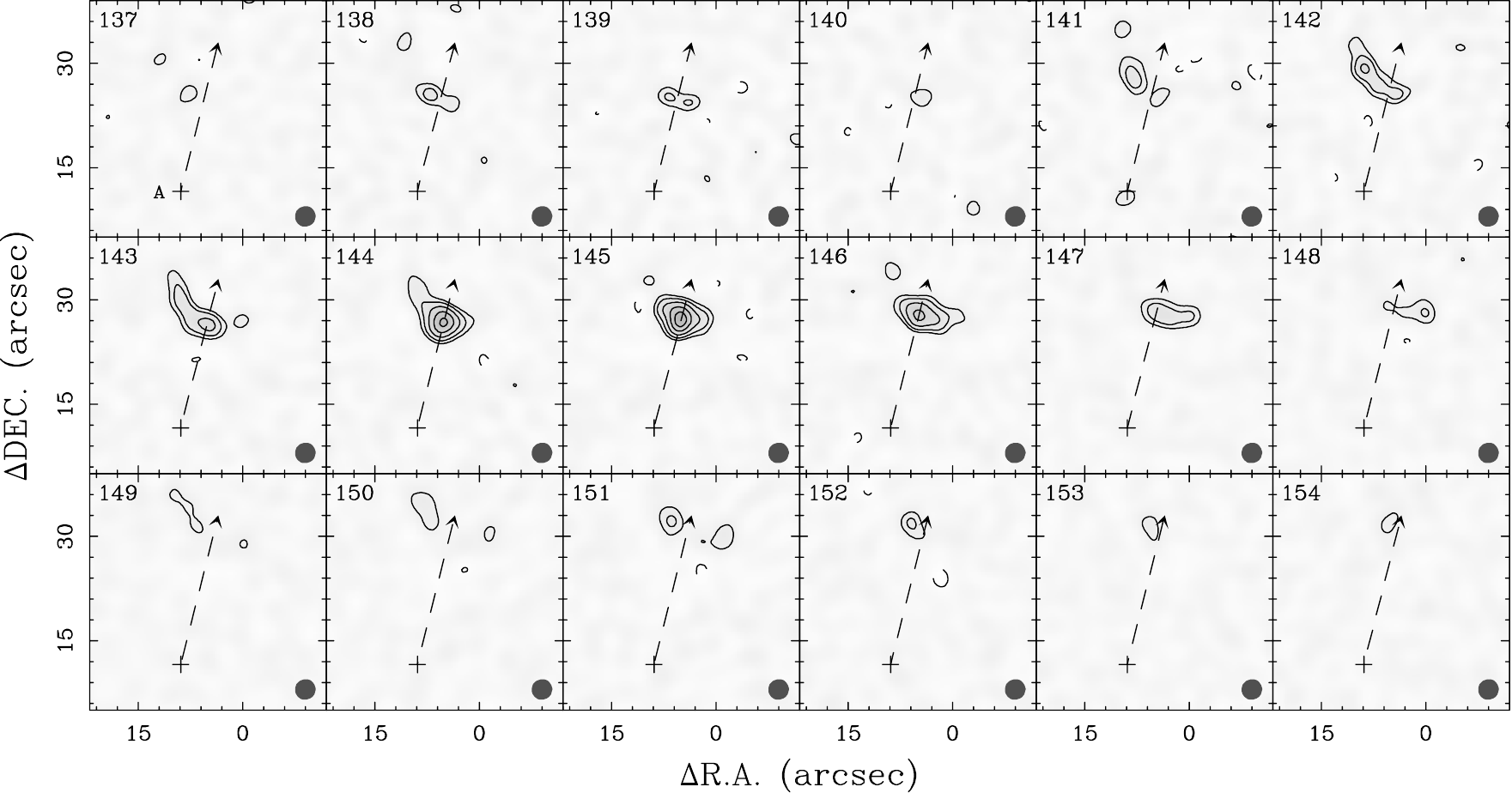}
\caption{SMA CO\,(2--1) velocity channel map of red bullet. Contours levels correspond to -3, 3, 5, 8, 12, and 16\,$\sigma$, where the 1\,$\sigma$ 
level is $\sim$\,30\,mJy beam$^{-1}$. In each panel, the cross marks the peak position of the dust continuum source SVS\,13\,A, while the dashed line 
arrow shows the direction of the redshifted jet.\label{channel_map_redbullet}}
\end{center}
\end{figure*}

\begin{figure*}
\begin{center}
\figurenum{10}
\includegraphics[width=16cm,angle=0]{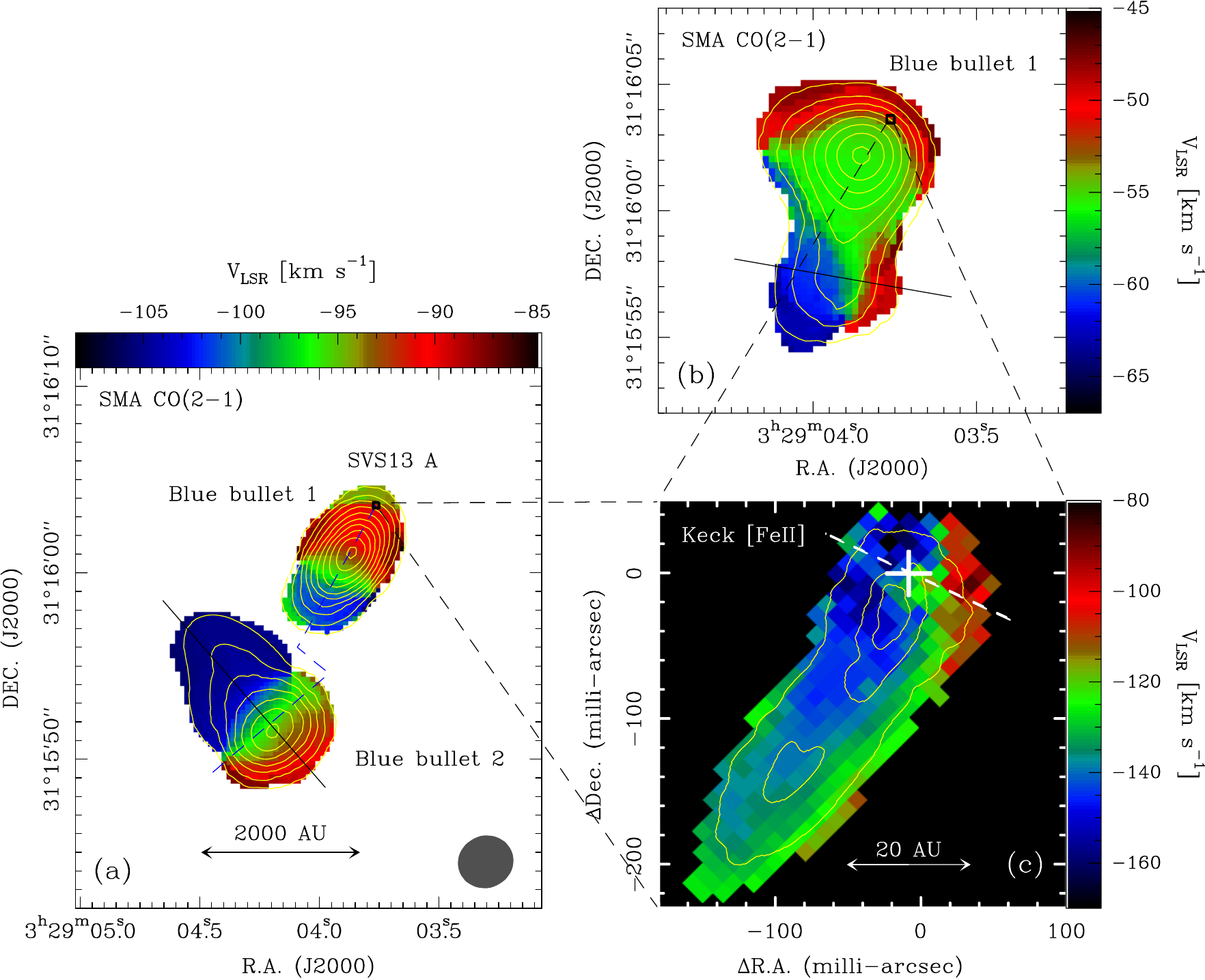}
\caption{(a) SMA CO\,(2--1) mean velocity field of blue bullets 2 and 1 (1st moment map; from Figure~2). (b) SMA CO\,(2--1) mean 
velocity field of the low velocity part of blue bullet 1 (1st moment map; from Figure~5). (c) Velocity map of the continuum-subtracted 
[Fe II] 1.644\,$\mu$m data of the SVS\,13 microjet, taken with the Keck telescope in 2012 (from Hodapp \& Chini 2014). The cross 
marks the position of the driving source SVS\,13 VLA\,4B. The white dashed line shows the direction of a tentative velocity gradient 
across the driving source.\label{rotation_keck}}
\end{center}
\end{figure*}

\begin{figure*}
\begin{center}
\figurenum{11}
\includegraphics[width=15cm,angle=0]{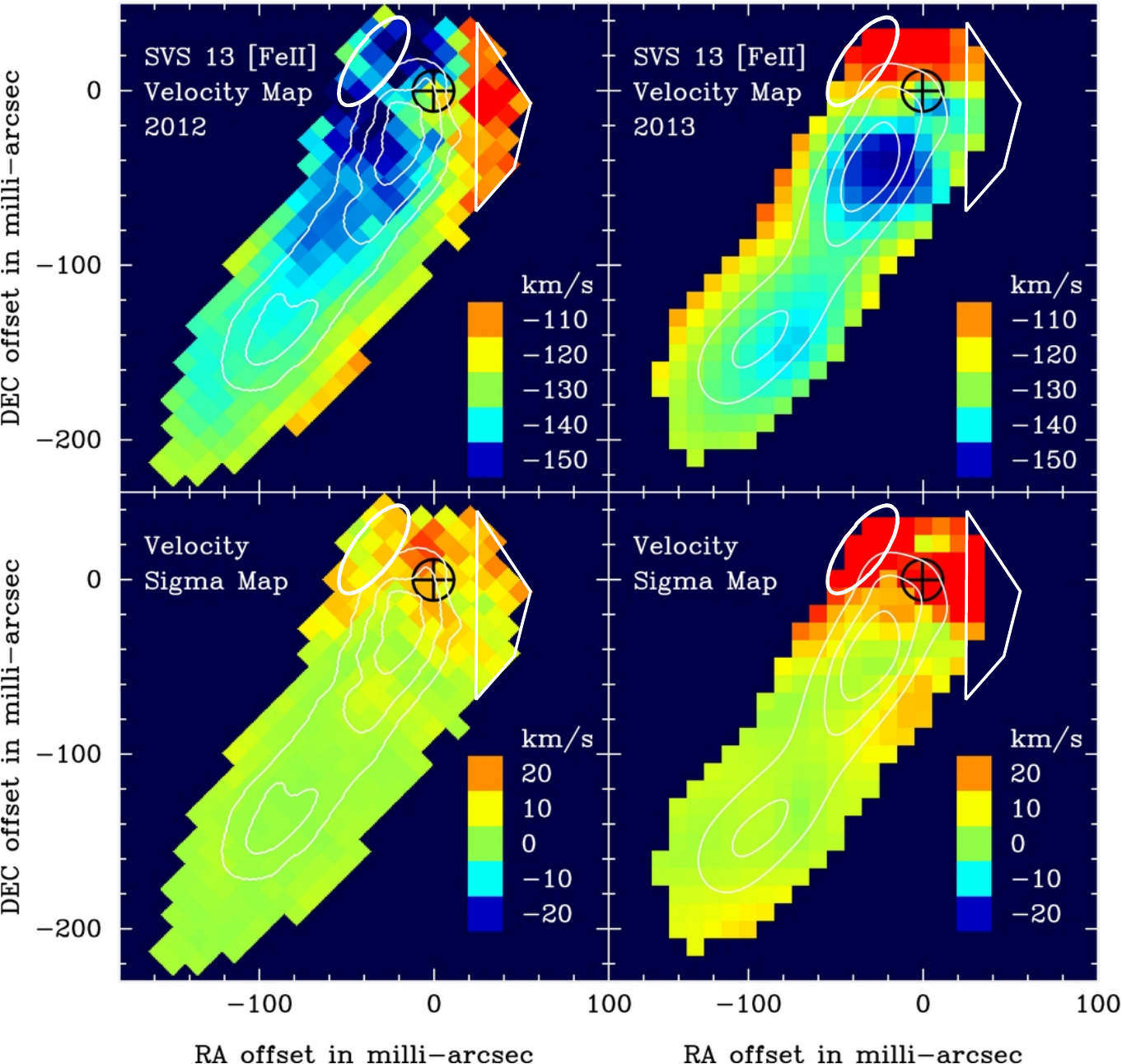}
\caption{The image is adapted from Figure~9 in Hodapp \& Chini (2014). Top panels: velocity map of the continuum-subtracted [Fe~II] 
1.644\,$\mu$m wavelength data cubes of the SVS\,13 jet, observed in 2012 and 2013, respectively. The bottom panels show the 
corresponding velocity sigma maps, showing the rms variations of the velocity measurements on the individual data cubes. The white 
ellipse and trapezium in each panel mark the same places where a tentative velocity gradient appears in the 2012 data.\label{keck2}}
\end{center}
\end{figure*}


\end{document}